\documentclass[11pt]{article}

\usepackage{graphicx}
\usepackage{epstopdf}
\usepackage{amsmath}
\usepackage{amssymb}
\usepackage{amsfonts}
\usepackage{amsthm}
\usepackage[usenames]{color}
\usepackage{array}
\usepackage{mathrsfs}
\usepackage{epic,eepic,pstricks}
\usepackage{rotating}

\usepackage[
      colorlinks=true,
      linkcolor=blue,
      urlcolor=blue,
      filecolor=blue,
      citecolor=red,
      pdfstartview=FitV,
      pdftitle={},
      pdfauthor={},
      pdfsubject={},
      pdfkeywords={},
      pdfpagemode=None,
      bookmarksopen=true
]{hyperref}

\usepackage{epsfig}

\usepackage{hyperref}

\begin{document}
\begin{flushright}
\texttt{\today}
\end{flushright}

\begin{centering}

\vspace{2cm}

\textbf{\Large{
Tachyon Inflation in Teleparallel Gravity}}

  \vspace{0.8cm}

 {\large  Amin Rezaei Akbarieh$^{1,2}$, Yousef Izadi$^3$ }

  \vspace{0.5cm}
  
\begin{minipage}{.9\textwidth}\small
\begin{center}
{\it $^1$ Department of Theoretical Physics and Astrophysics, University of Tabriz,1666-16471, Tabriz, Iran. }\\

{\it $^2$ Research Institute for Astronomy and Astrophysics of Maragha (RIAAM),
Maragha, Iran, P. O. Box: 55134 - 441.  }\\

{\it $^3$ Department of Physics, Kansas  State  University, 116  Cardwell  Hall,\ Manhattan, KS 66506, USA.}

\vspace{0.5cm}

{\tt am.rezaei@tabrizu.ac.ir, izadi@ksu.edu}
\\ $ \, $ \\

\end{center}
\end{minipage}
\begin{abstract}
We present a tachyonic field inflationary model in a teleparallel framework. We show that
tachyonic coupled with the f(T) gravity model can describe the inflation era in which f(T) is an arbitrary function of torsion scalar T. For this purpose, dynamical behavior of the tachyonic field in different potentials is studied, it is shown that the tachyonic field with these potentials can be an effective candidate for inflation. Then, we discuss slow-roll conditions and show that by the appropriate choice of the parameters, the inflation era can be explained via this model. Finally, we argue that our model not only satisfies the result of BICEP2, Keck Array and Plank for the upper limit of $r < .012$ but also, the obtained value for spectral index $n_{s}$ is compatible with the results of Plank and also Plank + WMAP + HighL + BAO at the 68\% confidence level.
\end{abstract}
\end{centering}
\section{Introduction}
According to the observations of WMAP \cite{Kinney:2003uw} and Plank \cite{Ade:2015tva}, the inflation era in the early universe is one of the most important parts of standard cosmology. The idea of inflation not only solves the flatness, horizon, magnetic monopoles problem but also predicts adiabatic and scale invariant Gaussian perturbation, responsible for the structure formation. This prediction is well consistent with the observational results \cite{Dvali:2003vv}.

Inflationary theories are formed based on the dynamics of an inflaton field which is minimally coupled with Einstein's gravity. Fundamental theories like supergravity, string theory, etc. suggest various sorts of potentials for inflation which are determined from phenomenological reasons or requirement of satisfaction of specific symmetries. Recent observations of Plank \cite{Ade:2015lrj} and BICEP \cite{Ade:2014xna} have made serious constraints on inflationary models, even some of the models have been discarded. Many inflationary models have recently been extensively studied \cite{Lyth:2007qh}. In the old inflationary model \cite{Pilo:2004mg}, it is assumed that inflaton is in the semi-stable false vacuum. It decays to the real vacuum through the first kind phase transition. This model is not able to describe a proper graceful exit.
Nowdays, many different inflationary models such as chaotic \cite{Linde}, extended \cite{Casas:1991zf}, power-law \cite{Feinstein:2002aj}, hybrid \cite{Linde:1993cn}, natural \cite{delaFuente:2014aca}, supernatural \cite{Randall:1996ip}, extranatural \cite{ArkaniHamed:2003wu}, eternal \cite{Guth:2007ng},brane   \cite{Burgess:2001fx}, oscillating \cite{Lee:1999pta}, trace-anomaly driven \cite{Bamba:2014jia}, ghost \cite{ArkaniHamed:2003uz}, tachyonic \cite{Sami:2002fs} and etc \cite{Ferraro:2006jd} have been proposed in the literature.

 One of the interesting methods for creating an inflationary model is to use tachyonic scalar field. In string theory, in addition to the non-Bogomol’nyi-Prasad-Sommerfield(BPS)D-branes, there are some other unstable D-branes which are known as non-BPS D-branes. These unstable D-branes are specified by having a single mode with negative mass \cite{Sen}. There are some terms in the potential of a tachyonic field which cause it to have a lower bound. Thus, non-BPS D-branes can decay to the minimum state of the tachyonic potential. Therefore, one can see that in the stable point, the sum of negative and minimal energy of tachyonic potential and the positive energy of the branes's tension is exactly zero \cite{Stable}. As a result, the unstable non-BPS branes, located in a flat-spacetime vacuum should decay into a real vacuum. So, the tachyonic field can be used to construct an inflationary model. It can also be shown that the tachyonic field can describe the accelerated expansion of the universe. Therefore, in addition to the explanation of the inflationary era, tachyonic models are suitable candidates for dark energy \cite{Motavalli}.

 In 1993, Raiten showed that the tachyonic potential could be an appropriate solution for the inflationary era \cite{Raiten:1993pn}. Although some people were interested in the tachyonic model, in a paper which was published in 2002 \cite{Kofman:2002rh} Kofman and Linde showed that is it not possible to explain the inflationary era using the simple tachyonic model. Usually, the inflation could only accrue in super-Planckian densities, where the four-dimensional effective field theories cannot be used. Since,  in these models, the tachyonic field does not oscillate in the minimum of the potential, creation of matter and reheating face problem. They also claimed that the tachyonic field does not play any role after the inflation era. Surely, these problems do not arise in hybrid inflation models. In \cite{Sami:2002fs}, details of tachyonic inflation have been investigated with regard to the exponential potential and with the analyze of phase space diagram of the tachyonic model, it was shown that the dust like answer is an absorbent. They also concluded that the reheating in the tachyonic model is problematic. Unlike the two previous papers, it has been shown in \cite{Li:2002et} that rolling of tachyon could be a suitable source for the inflation era. In \cite{Piao:2002nh}, a tachyonic model with a non-minimal coupling with gravity has been studied. It has been shown that, with a particular coupling between tachyon and gravity, the model is compatible with the observational results and could resolve various existing problems in mono and multi tachyonic models. In \cite{Steer:2003yu}, tachyonic inflation model is compared with the standard scalar method, it is shown that in low order perturbation, both models reach similar results. This paper also discusses the compatibility of some of the tachyonic potentials with the observations as the results agrees with the observations at the limit of $\sigma =1$. All these models predict a very small and negative value for the running of the scalar spectral index.
In 2006, Cardenas \cite{Cardenas:2006py} presented a valuable tachyonic quintessential inflationary model to explain the period of inflation in the early universe as well as the current accelerated expansion. This paper claimed that the reheating stage plays an essential role in having a compatible cosmology and obtaining an accelerating period. In \cite{Herrera:2008bj} tachyonic inflation in curved spacetime is investigated, and exact solutions for the field, pressure, scalar factor and some of the other cosmological parameters are obtained. Tachyon Logamediate inflation model is studied in \cite{Ravanpak:2015bta} and \cite{Kamali:2017nwe}. In \cite{Nozari:2014qba} the compatibility of the cosmic inflation model with the Plank + BAO + BICEP2 + WMAP has been discussed. While the tachyonic field inflation with the minimal coupling is consistent with the data of Plank 2013, it has not been confirmed by Plank + WMAP+ BICEP2 +BAO. Nevertheless, the non-minimal model is compatible with this data set. Since the non-minimal tachyonic field is fitting better to the observational data, we decided to present a non-minimal tachyonic model in teleparallel gravity.

 In this paper, using a tachyonic field in teleparallel gravity, we propose an inflationary model.  Teleparallel gravity is an attempt made by  Einstein \cite{Cai:2015emx} to establish a unified theory of gravity and electromagnetic based on the mathematical structure of the teleparallel framework. General relativity is a gauge theory of the gravitational field based on the equivalence principle. Although it is not necessary to work with the Riemannian manifold, there are several modified theories such as Riemann-Cartan in which the geometrical structure of the theory is not metrical. There are more than one dynamic variables in these kinds of modifications \cite{L. Smalley}. If we neglect the non-metrical nature of the theory, we can move from Riemannian manifold to Weitznbock spacetime with local Riemannian tensor and torsion. One example of these sort of theories is teleparallel gravity in which we work with the non-Riemannian manifold. In the teleparallel framework, the dynamics of the metric is determined by scalar torsion $T$. Basic variables in teleparallel gravity are four Vierbein basis $e_{\mu}^{i}$ which are orthogonal. Free coordinate basis are given by
\begin{equation}
g_{\mu\nu}=\eta_{ij} e_{\mu}^{i}e_{\nu}^{j}.
\end{equation}
 These four basis are orthonormal while $\eta_{\mu\nu}$ is Minkowski metric, it means that $e_{\mu}^{i}e_{j}^{\mu}=\delta_{i}^{j}$. This is the simplest kind of teleparallel gravity which is known as $f(T)$ gravity, where $f(T)$ is an arbitrary function of torsion T. Some authors believe that curvature and torsion could have a different role in the early universe. In fact, many reasons suggest that the torsion and curvature naturally produce the repulsive portions to the energy-momentum tensor \cite{ Sabbata}. Much evidence indicates that the torsion might be involved in any comprehensive theory which works with non-gravitational fundamental interactions \cite{Capozziello}. The torsion is usually interpreted as a spin density of the matter, but often it has a more general and broader meaning.

In this paper, in addition to the introduction of an inflationary model based on teleparallel gravity, the phenomenological aspects of this model will also be considered. Using the description of phase space, we will study the dynamical behavior of the tachyonic inflationary model. This paper is organized as follows. In the second section, we present a short introduction to teleparallel gravity, and we study properties of the dynamical behavior of the tachyonic model in a teleparallel framework. The third section is devoted to the study of a tachyonic inflationary model. In section four we present phenomenological results of the model. In the final section, a conclusion is given.
\section{Dynamical behavior of tachyonic field in $f(T)$ gravity}
In terms of teleparallel gravity, vierbein $e_{\nu}^{i}$ are the main quantity. As mentioned, these basis are orthonormal and independent of the coordinates system and are defined in terms of the metric as follows
 \begin{equation*}
 g_{\mu\nu}=\eta_{ij} e_{\mu}^{i}e_{\nu}^{j}.
 \end{equation*}

In which $\mu$ and $\nu$ are coordinates on the specified curve and take $0,1,2,3$. In teleparallel gravity, Weizenbock connection tensors are given by
\begin{eqnarray}
\Gamma_{\mu\nu}^{\alpha}= e_{i}^{\alpha}\partial_{\nu}e_{\mu}^{i}=-e_{\mu}^{i}\partial_{\nu} e_{i}^{\alpha}.
\end{eqnarray}
Components of the torsion tensor are obtained from anti-symmetric part of Weitznbock connection
\begin{equation}
\Gamma_{\mu\nu}^{\alpha}=\Gamma_{\nu\mu}^{\alpha}-\Gamma_{\mu\nu}^{\alpha}=e_{i}^{\alpha}\left(\partial_{\mu} e_{\nu}^{i}-\partial_{\nu} e_{\mu}^{i}\right).
\end{equation}
Also, torsion  scalar is defined as
\begin{equation}
T=T_{\mu\nu}^{\alpha}S_{\alpha}^{\mu\nu},
\end{equation}
in which we have a new tensor $S_{\alpha}^{\mu\nu}$ defined as
\begin{equation}
S_{\alpha}^{\mu\nu}=\frac{1}{2}\left(K_{\alpha}^{\mu\nu}+\delta_{\alpha}^{\mu}T_{\beta}^{\beta\nu}-\delta_{\alpha}^{\nu}T_{\beta}^{\beta\mu}\right).
\end{equation}
Now we can write the action for the tachyonic field in teleparallel gravity as \cite{Banijamali:2012nx}
\begin{equation}
S=\int d^4x \sqrt{-g}\left\{f(\phi)T-V(\phi)\sqrt{1+g^{\mu\nu} \partial_{\alpha}\phi   \partial_{\beta}\phi}\right\},
\end{equation}
where $\phi$ is a scalar field which is non-minimally coupled to torsion scalar $T$ with a coefficient of  $f(\phi)$. $V(\phi)$ represents the potential function for the scalar field which can be determined using the physical conditions and symmetric considerations. The introduced action with torsion formalism in general relativity is similar to the standard scalar-tensor gravity in which the scalar field is coupled to the Ricci Scalar \cite{Y.-F. Cai}.
In this paper, we assume that the background geometry of the universe is flat and described by Friedmann-Robertson-Walker (FRW) metric
\begin{equation}
ds^2=-dt^2+a(t)^2\left(dx^2+dy^2+dz^2\right),
\end{equation}
where $a(t)$ is the scalar factor. Using equation (2.4), one can easily calculate the torsion scalar in flat FRW metric
\begin{equation}
T=-6H^2,
\end{equation}
In which $H=\dot{a}/a$ is the Hubble parameter, and the dot denotes differentiation with respect to the time. In this model, it is assumed that the scalar field $\phi$ is only a function of time. Using torsion scalar (2.8), we can show that point-like Lagrangian density for action (2.6) is obtained as

\begin{eqnarray}
\mathcal{L}=-6 f(\phi)a \dot{a}^2-a^3 V(\phi)\sqrt{1-\dot{\phi}^2}.
\end{eqnarray}
Euler-Lagrange equation for a(t) and $\phi(t)$ are obtained respectively as
\begin{equation}
\frac{\ddot{a}}{a}=-\frac{1}{2} H^2-\frac{f'(\phi)}{f(\phi)}\dot{\phi}H+\frac{1}{4} \frac{V(\phi)}{f(\phi)}\sqrt{1-\dot{\phi}^2},
\end{equation}
\begin{equation}
\ddot{\phi}+\left(1-\dot{\phi}^2\right)\left\{\sqrt{1-\dot{\phi}^2} \frac{6f'(\phi)}{V(\phi)}H^2+\frac{V'(\phi)}{V(\phi)}+3H\dot{\phi}\right\}=0.
\end{equation}
According to the equation $T_{\mu\nu}=-\frac{2}{\sqrt{-g}}\frac{\delta S}{\delta g_{\mu\nu}}$, one can calculate the energy momentum tensor. Therefore, energy density and pressure of tachyonic field can be written as
\begin{equation}
\rho_{\phi}=\frac{1}{16 \pi G f(\phi)}\frac{V(\phi)}{\sqrt{1-\dot{\phi}^2}},
\end{equation}
\begin{equation}
P_{\phi}=\frac{1}{8\pi G}\left\{H^2+2\frac{f'(\phi)}{f(\phi)}\dot{\phi}H+\frac{1}{6} \frac{V(\phi)}{f(\phi)}\frac{3\dot{\phi}^2-4}{\sqrt{1-\dot{\phi}^2}}\right\}.
\end{equation}
If we assume that $\dot{\phi}$ is a function of $\phi$, we can rewrite the second derivative $\ddot{\phi}$ as
\begin{equation}
\ddot{\phi}=\dot{\phi}'(\phi)\dot{\phi},
\end{equation}
In which prime indicates differentiation with respect to $\phi$. Considering Friedmann equation $H^2=\frac{8\pi G}{3}\rho $ along with equation (2.14), equation (2.11) turns out to be
\begin{equation}
{\dot{\phi'}}\dot{\phi}+\left(1-\dot{\phi}^2\right)\left\{\sqrt{1-\dot{\phi}^2} \frac{6f'(\phi)}{V(\phi)}(\frac{8\pi G}{3}\rho)+\frac{V'(\phi)}{V(\phi)}+3\sqrt{\frac{8\pi G}{3} \rho}\dot{\phi}\right\} =0.
\end{equation}
Now, if we substitute tachyonic energy density in equation (2.15), then we have
\begin{equation}
\frac{d\dot{\phi}}{d\phi}= \frac{\dot{\phi}^2-1}{\dot{\phi}}\left\{\frac{f'(\phi)V(\phi)}{f^2(\phi)}+\frac{V'(\phi)}{V(\phi)}+\left(\frac{3}{2}\frac{V(\phi)\dot{\phi}^2}{f(\phi)\sqrt{1-\dot{\phi}^2}}\right)^{\frac{1}{2}}\right\}.
\end{equation}
Now, for the different modes of coupling coefficient of the tachyonic field to teleparallel gravity, we draw a phase space diagram and give a physical interpretation. For each of coupling coefficient, various potentials will be considered.

\textbf{I)\boldmath$f(\phi)=1$}: For a state in which there is no coupling between tachyonic field and torsion scalar, or $f(\phi)=1$( or any constant value), we can see that equation (2.16) is simplified 

\begin{equation}
\frac{d\dot{\phi}}{d\phi}= \frac{\dot{\phi}^2-1}{\dot{\phi}}\left\lbrace\frac{V'(\phi)}{V(\phi)}+\left(\frac{3}{2}\frac{V(\phi)\dot{\phi}^2}{\sqrt{1-\dot{\phi}^2}}\right)^\frac{1}{2}\right\rbrace.
\end{equation}
Tachyonic potential in the open string theory \cite{Steer:2003yu} is given by
\begin{equation}
V(\phi)=\frac{V_{.}}{cosh(\frac{\phi}{{\phi}_{.}})},
\end{equation}
Where ${\phi}_{.}$ is equal to $\sqrt{2}$ for non-BPS brane in superstring theory and $2$ in Bosonic string theory. Note that tachyonic field at $\phi \rightarrow \infty $ has a ground state. Substituting this potential in equation (2.18), plot of the phase space diagram $\dot{\phi} - {\phi}$ is obtained as figure (1).
\begin{figure}[h!]
\centering
\includegraphics[width=8cm]{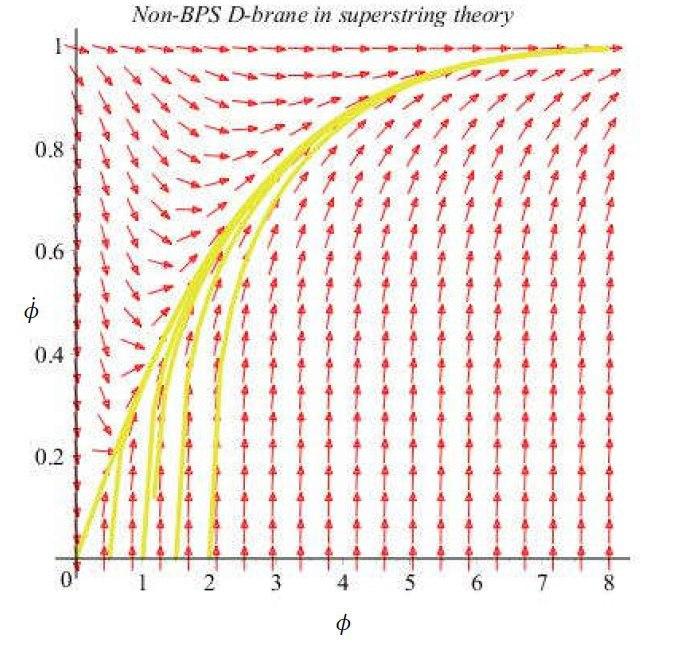}
\caption{Phase space diagram for $V(\phi)=\frac{V_{.}}{cosh(\frac{\phi}{{\phi}_{.}})}$ with $V_{.}=\frac{2}{3}$, $\phi_{.}=\sqrt{2}$ and coupling constant $f(\phi)=1$}
\label{figure 1}
\end{figure}\\
Note that by increasing the field value (with arbitrary boundary conditions) $\dot{\phi}$ tends to one. So, the term for the scalar field vanishes in action (2.6). In this case, it can be seen that the equation of state is valid for the condition of accelerated expansion, $\omega=\frac{p}{\rho}\leqslant\frac{-1}{3}$.\\
The tachyonic potential in anti-D brane theory appears as $V(\phi)=V_{.}e^{\frac{1}{2}m^2\phi^2}$, which is the excited state for massive scalar fields. This potential has a minimum at $\phi=0$. Now we substitute this potential in equation (2.17) to get the following phase space diagram.
\begin{figure}[h!]
\centering
\includegraphics[width=8cm]{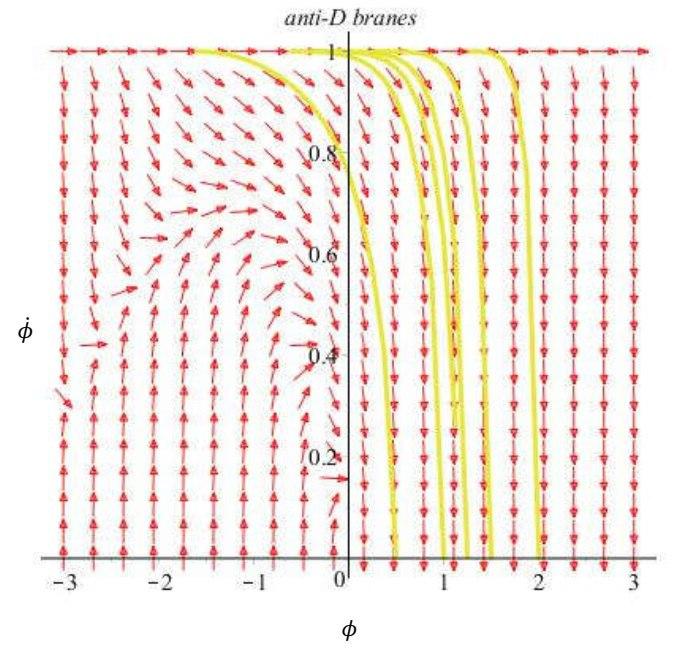}
\caption{Phase space diagram for $V(\phi)=V_{.}e^{\frac{1}{2}m^2\phi^2}$ with $V_{.}=\frac{2}{3}$,$m=1$ and coupling constant $f(\phi)=1$}
\label{figure 2}
\end{figure}
For this potential, it can be seen that by decreasing the field value, the derivative of the field approaches to one. So, the tachyonic term in action (2.6) tends to zero. For such a potential in the state of decreasing value of the field, we expect that this model results in an accelerated expanding period. Other potentials can be considered for the tachyonic field. Some of these potentials are obtained by imposing Noether symmetry for the action (16), like $V(\phi)=V_{.}e^{-\frac{\phi}{\phi_{.}}}$

It has been shown in \cite{Sen} that taking into account this potential, the solutions of the tachyonic field will satisfy Noether symmetry in a teleparallel framework. Unlike the two previous potentials which were even functions of the tachyonic field, this potential is decreasing with respect to $\phi$ and with the increment of the tachyonic field, its effects is eliminated from the action (2.6). Now we can draw a phase space diagram of this potential as follows.\\
\begin{figure}[h!]
\centering
\includegraphics[width=8cm]{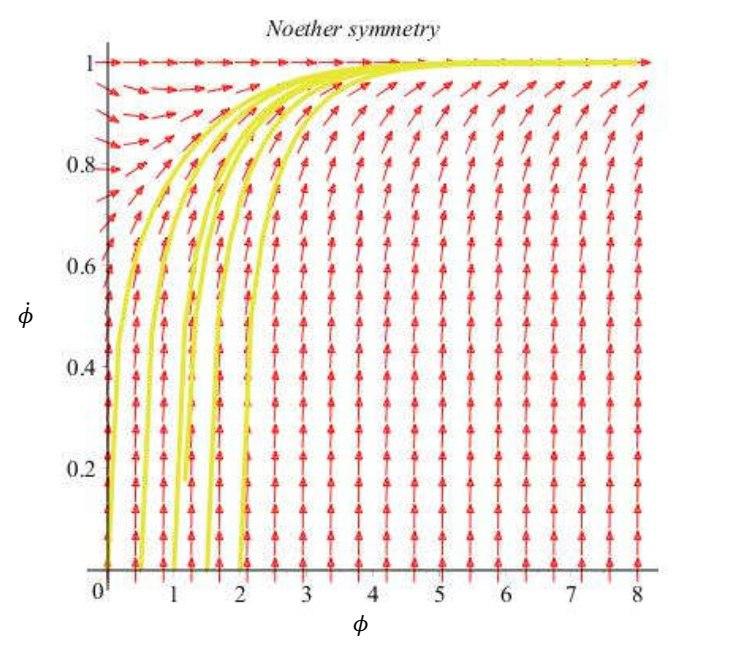}
\caption{Phase space diagram for $V(\phi)=V_{.}e^{-\frac{\phi}{\phi_{.}}}$ with $V_{.}=\frac{2}{3}$,$\phi_{.}=0$ and coupling constant $f(\phi)=1$}
\label{figure 3}
\end{figure}
It can be seen from figure (3) that with the increment of the field value, both the potential and the term $\sqrt{1-\dot{\phi}^2}$ simultaneously tend to zero. Therefore, as tachyonic potential in non-BPS D-brane in superstring theory, with the increasing the field value, we expect a positive acceleration period in this potential. We can also point out other potentials which are inverse power, $ V(\phi)\varpropto \phi^{-n}$. Figure (4) is the phase space diagram for this type of potentials.
\begin{figure}[h!]
\centering
\includegraphics[width=8cm]{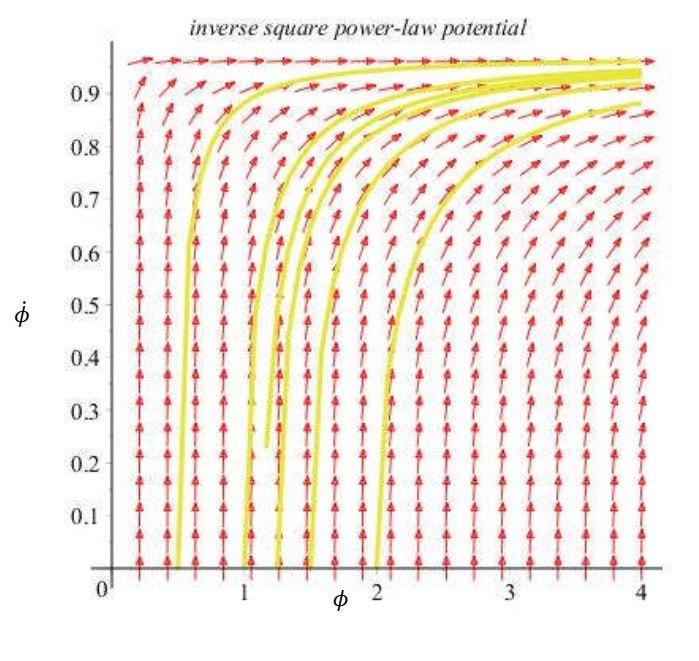}
\caption{Phase space diagram for $V(\phi)= V_{.}\phi^{-2}$ with $ V_{.}=\frac{2}{3}$ and coupling constant $f(\phi)=1$}
\label{figure 4}
\end{figure}
\begin{figure}[h!]
\centering
\includegraphics[scale=.45]{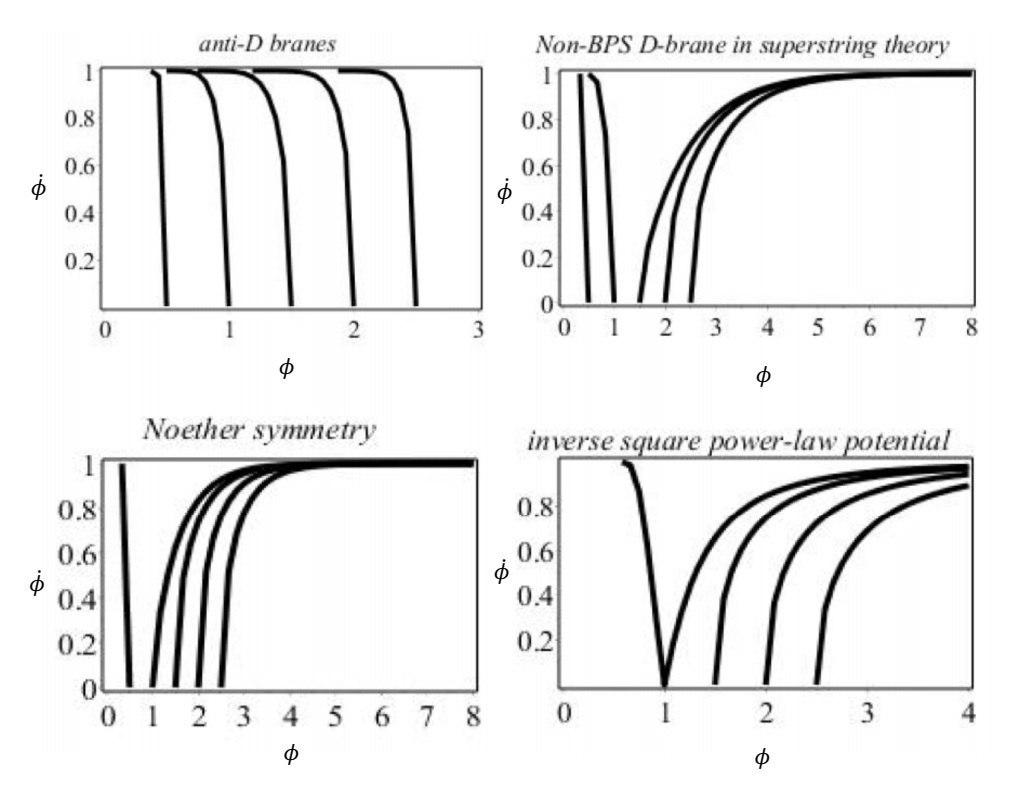}
\caption{Phase space diagram for the different potentials with coupling constant $f(\phi)=\phi^{2}$}
\label{figure 5}
\end{figure}
As we can see, the results of the dynamical behavior of these potentials are similar to what obtained for the potentials in non-BPS D-brane in superstring as well as the potentials found from Noether symmetry.\\
\textbf{II)\boldmath$f(\phi)=\phi^{2}$}: In this case we choose $f(\phi)=\phi^{2}$ for coupling of field with torsion scalar.
 For the different potentials used previously, we have the phase space diagrams depicted in figure (5).
As illustrated above, for $f(\phi)=\phi^2$ the dynamical behavior of the tachyonic field does not change for the used potentials. The only difference is that except anti-D brane, we can see in the other potentials that for the boundary condition $\phi<1$, decreasing of the tachyonic field, causes $\dot{\phi}$ tending to one and tachyonic terms will be decoupled from teleparallel gravity.
\section{Cosmological inflation in teleparallel gravity}
Recently, many studies have been carried out on the inflation model of tachyonic fields \cite{Steer:2003yu}. It has also been shown that the tachyonic inflation model suffers from serious problems \cite{Kofman:2002rh}. However, the authors of paper \cite{Piao:2002nh} succeeded in solving these problems by presenting a multi-tachyonic inflation model. Throughout this paper, it is assumed that there is a(non)minimum coupling between the tachyon field and gravity \cite{Sen:2003mv}. However, In this paper, we study a novel model in which the tachyonic field is coupled to the teleparallel gravity. In this section, we consider the case in which $f(\phi)=1$. Note that when the coupling is non-minimal, conformal transformation can be used. Conformal transformation in $f(T)$ gravity has been studied in paper\cite{Yang:2010ji}. Unlike to conformal transformation in $f(R)$ gravity, a term which is related to the coupling of scalar and tensor $\dot{\phi}T_{\rho0}^{\rho}$ appears.
Following the method in \cite{Yang:2010ji}, if we apply the conformal transformation $g_{\mu\nu}(x)\rightarrow f(\phi)g_{\mu\nu}$ on action (2.6), we obtain
\begin{equation}
S_{EF}=\int d^{4}x \sqrt{-g}\left\{T+3\frac{f'(\phi)}{f^3{(\phi)}}\dot{\phi}T_{\rho0}^{\rho}-3 f\frac{f'^{2}(\phi)}{f^{4}(\phi)}\dot{\phi}^2\right\}+
\int d^4x\sqrt{-g}\frac{V(\phi)}{f^2(\phi)}\sqrt{1-f^2(\phi)\dot{\phi}^2},
\end{equation}

in which $T_{\rho}^{\rho 0}=\frac{51}{2}H$. The first condition that needs to be satisfied is, $\ddot{a}>0$. Thus, from equation (2.10), we have
\begin{equation}
-\frac{1}{2}H^{2}-\frac{f'(\phi)}{f(\phi)}\dot{\phi}H+\frac{1}{4}\frac{V(\phi)}{f(\phi)}\sqrt{1-\dot{\phi}^{2}}>0.
\end{equation}
Since we only study the minimal coupling case, if we assume $f(\phi)=1$ , we find
\begin{equation}
-\frac{1}{2}H^2+\frac{1}{4}V(\phi)\sqrt{1-\dot{\phi}^2}>0.
\end{equation}
By inserting $H^2=\frac{8\pi G}{3}\rho$ and using equation (2.12), we obtain
\begin{equation}
-\frac{1}{\sqrt{1-\dot{\phi}^{2}}}+3\sqrt{1-\dot{\phi}^{2}}>0.
\end{equation}
From the above equation, we can conclude that $\dot{\phi}^2(t)<\frac{2}{3}$. To describe an appropriate inflation period, the tachyonic field of this model should start from a very small initial value of $\dot{\phi}$. Now, the slow-roll parameters are computed as
\begin{equation}
\epsilon=\frac{1}{2} M_{p}^{2}\left(\frac{V'(\phi)}{V(\phi)}\right)^2, \quad \quad \eta=M_{p}^{2}\left(\frac{V''(\phi)}{V(\phi)}\right),
\end{equation}
where $\epsilon$ and $\eta$ are the slow-roll parameters. The number of $e$-folds is given by
\begin{equation}
N=\int\frac{1}{\sqrt{2\epsilon}}\frac{d\phi}{M_{p}^2}.
\end{equation}
Spectral index $n_{s}$ and the tensor to the scalar ratio $r$ are given by
\begin{equation}
n_{s}=1-6\epsilon+2\eta, \quad \quad r=16\epsilon.
\end{equation}
For potential of open string theory, the above parameters are found to be
\begin{equation}
\epsilon\approx \frac{1}{2}M_{p}^2 \frac{1}{\phi_{0}^2}tanh^2\left(\frac{\phi}{\phi_{0}}\right)<<1, \quad
\quad \eta\approx M_{p}^2 \frac{1}{\phi_{0}^2}\left(2 tanh^2\left(\frac{\phi}{\phi_{0}}\right)-1\right)<<1.
\end{equation}
If we choose $\frac{\phi}{\phi_{0}}$ small enough, the condition (3.26) is guaranteed to be satisfied.
Then, by integrating equation (3.24), the number of $e$-folds is calculated as
\begin{equation}
N=\frac{\phi_{0}^2}{M_{p}}Ln\left(Sinh(\frac{\phi}{\phi_{0}})\right)\biggr\rvert_{\phi_{end}}^{\phi_{N}}.
\end{equation}
Since $\epsilon=1$ at the end of inflation era, one can find
\begin{equation}
\phi_{end}=\phi_{0} Arctanh\left(\sqrt{2}\frac{\phi}{M_{p}}\right).
\end{equation}
According to the observations of the cosmic background radiation, $N$ is approximately 57.7 \cite{Bezrukov:2013fka}, so $\phi_{N}$ can be calculated. If we take $\phi_{0}\sim 0.1$, we find $\phi_{N}\approx 500 M_{p}$. Considering the results in \cite{Ade:2013zuv}, the amplitude of scalar power spectrum is restricted to
\begin{equation}
\Delta_{R}^2=\frac{1}{8\pi^{2}}\frac{H^2}{\epsilon M_{p}^2}\approx 2* 10^{-10}.
\end{equation}
This restriction could be used to determine the value of $\phi_{0}$. Using the equations for the slow-roll region, one can state that $\frac{V(\phi)}{\epsilon}\approx (0.03 M_{p}^4)$. Since $N\approx 57.7$, so we find spectral index to be $n_{s}\approx 0.956$ and the tensor to the scalar ratio $r\approx 0.0061$. Although there are a lot of uncertainty about the results of BICEP2 which reports $r\approx 0.2$ \cite{Ade:2014xna}, Keck Array has confirmed their results in the excess of B-mode power over the standard expectation \cite{Ade:2015fwj,Ade:2015tva}.
 If we look at the results of BICEP2, Keck Array and Plank all together, we deduce that $r<0.12$ and the likelihood maximum is around $r\approx 0.05$ \cite{Ade:2015xua}. Even though the obtained value for $r$ from the tachyonic model is different from the maximum value, it satisfies the upper limit in \cite{Ade:2015xua}.
\\

\section{Conclusion}
In this paper, we examined the tachyonic field in the framework of teleparallel gravity. After introducing the model, we discussed the dynamical behavior of the tachyonic field under the influence of different proposed potentials. For the case in which there is no coupling between tachyonic and the scalar field ($(f(\phi)=1$ or any constant), we argued that for tachyonic field in open string theory and D-brane theory and also tachyonic potentials consistent with Noether's theorem, tachyonic field is a proper candidate for inflation.
For the tachyonic potential in open string theory with increasing the field's value (with an arbitrary boundary condition), $\dot{\phi}$ approaches to one. Thus, the scalar field term in action for the tachyonic field is eliminated. Therefore, it is straightforward to see that the equation of state satisfies the accelerating expansion period condition $\omega=\frac{p}{\rho} \leqslant -\frac{1}{3}$.

For the tachyonic potential of an anti-D-brane, it can be concluded that in the case of a tachyonic field reduction, we expect that this model leads to a period with accelerated expansion. By examining the dynamical behavior of tachyonic field in the potential derived from Noether symmetry, one can see that with increasing the field value, both potential and the conjunction term $\sqrt{1-\dot{\phi}^2}$ simultaneously approach to zero. Therefore, like the tachyonic potential in Non-BPS brane in superstring theory, with the increase in the field strength, we expect to have a positive acceleration for this potential.

Also, by studying the dynamical behavior of the tachyonic field in $f(\phi)=\phi^{2}$, we obtained a similar result. The only difference that we see for all potentials except for anti-D-brane is that for the boundary condition $\phi <1$, with reduction of the field value, $\dot{\phi}$ tends to zero and tachyonic terms are decoupled from gravitation.

Finally, we studied inflation of the tachyons in the framework of $f(T)$ and considered the state in which $f(\phi)=1$. Then with the computation of the slow-roll parameters for this model, we found that with the proper choice for the parameters, one can find approximately $n_{s}=0.956$ for the spectral index. Also, the tensor to the scalar ratio is obtained as $r=0.0061$. The calculated $r$ from tachyonic model agrees with the upper limited value obtained in \cite{Ade:2015xua}.

\section{Acknowledgement}
This work has been supported financially by Research Institute for Astronomy and Astrophysics
of Maragha (RIAAM) under research project No. 1/4165-67. We would like to thank Mahboub Hosseinpour and Praful Gagrani for their comments on the manuscript.

\end{document}